\documentclass[twocolumn,tighten,times]{aastex63}

\received{September 28, 2020}
\revised{October 26, 2020}
\accepted{October 28, 2020}
\submitjournal{ApJ}

\shorttitle{The pulsating white dwarf G117-B15A}
\shortauthors{Kepler et al.}

\newcommand{\Msun}{\ensuremath{M_\odot}}
\newcommand{\Pdot}{\ensuremath{\dot{P}}}

\begin{document}

\title{The pulsating white dwarf G117-B15A: still the most stable optical clock known}

\correspondingauthor{S.O. Kepler}
\email{kepler@if.ufrgs.br}

\author[0000-0002-7470-5703]{S. O. Kepler}
\affiliation{Instituto de F\'\i sica, Universidade Federal do Rio Grande do Sul, 91501-970 Porto Alegre RS, Brazil}
\author{D. E. Winget}
\affiliation{Department of Astronomy and McDonald Observatory, University of Texas at Austin, Austin TX 78712-1085, USA}
\author{Zachary P. Vanderbosch}
\affiliation{Department of Astronomy and McDonald Observatory, University of Texas at Austin, Austin TX 78712-1085, USA}
\author{Barbara Garcia Castanheira}
\affiliation{Baylor University, Department of Physics, Waco TX 76798, USA}
\author{J.J. Hermes}
\affiliation{Department of Astronomy,
Boston University,
Boston MA 02215, USA}
\author[0000-0002-0656-032X]{Keaton J. Bell}
\affiliation{DIRAC Institute, Department of Astronomy,
University of Washington,
Seattle WA 98195-1580, USA} 
\altaffiliation{NSF Astronomy and Astrophysics Postdoctoral Fellow}
\author{Fergal Mullally}
\affiliation{SETI Institute, 189 Bernardo Ave, Suite 200, Mountain View CA 94043, USA}
\author[0000-0002-0797-0507]{Alejandra D. Romero}
\affiliation{Instituto de F\'\i sica, Universidade Federal do Rio Grande do Sul, 91501-970 Porto Alegre RS, Brazil}
\author[0000-0002-6748-1748]{M. H. Montgomery}
\affiliation{Department of Astronomy and McDonald Observatory, University of Texas at Austin, Austin TX 78712-1085, USA}
\author{Steven DeGennaro}
\affiliation{Department of Astronomy and McDonald Observatory, University of Texas at Austin, Austin TX 78712-1085, USA}
\altaffiliation{IMDb}
\author{Karen I. Winget}
\affiliation{Department of Astronomy and McDonald Observatory, University of Texas at Austin, Austin TX 78712-1085, USA}
\author{Dean Chandler}
\affiliation{Meyer Observatory and Central Texas Astronomical Society, 209 Paintbrush, Waco TX 76705, USA,
chandler@vvm.com}
\author{Elizabeth J. Jeffery}
\affiliation{Physics Department, California Polytechnic State University, San Luis Obispo, CA 93407 USA, elizabeth.j.jeffery@gmail.com}
\author{Jamile K. Fritzen}
\affiliation{Instituto de F\'\i sica, Universidade Federal do Rio Grande do Sul, 91501-970 Porto Alegre RS, Brazil}
\author{Kurtis A. Williams}
\affiliation{Department of Physics and Astronomy, Texas A\&M University-Commerce, P.O. Box 3011, Commerce, TX, 75429-3011, USA}
\author{Paul Chote}
\affiliation{Department of Physics, University of Warwick, Coventry CV4 7AL, United Kingdom, p.chote@warwick.ac.uk}
\author[0000-0003-3609-382X]{Staszek Zola}
\affiliation{Astronomical Observatory of the Jagiellonian University, ul. Orla 171, PL-30-244 Kraków, Poland}
\affiliation{Mt. Suhora Observatory, Pedagogical University, ul. Podchorazych 2, PL-30-084 Kraków, Poland}


\begin{abstract}
The pulsating hydrogen atmosphere white dwarf star G~117-B15A has been observed
since 1974. Its main pulsation period at 215.19738823(63)~s, observed in optical light curves,  varies by only
$(5.12\pm 0.82)\times 10^{-15}$~s/s and shows no glitches, as pulsars do.
The observed rate of period change corresponds to a change of the pulsation period by 1~s in 6.2 million years.
We demonstrate that this exceptional optical clock can continue to put stringent limits on fundamental physics, such as constraints on interaction from hypothetical dark matter particles, as well as to search for the presence of external substellar companions.

\end{abstract}



\section{Introduction} \label{sec:intro}

\object[G 117-B15A]{G~117--B15A},
also called RY~LMi and WD~0921+354,
is a 
pulsating white dwarf with a hydrogen atmosphere,
a DAV or ZZ Ceti star
\citep{M79}.
White dwarf stars are the most common end product of stellar evolution. From the observed
initial-mass-function, more than 97\% of all stars evolve to white dwarfs 
\citep{Fontaine01,Koester02,Smartt09,Althaus10,Woosley15,Lauffer18}. When the normal white dwarf cooling reduces their temperatures such that
the their outer envelopes develop partial ionization zones --- which depends on the
dominant chemical element in the envelope --- convection zones are established that drive pulsations.
These pulsations are seen as luminosity variations and the period of the dominant pulsation mode is 
related to the thermal timescale at the base of the envelope.
These white dwarf stars
show multi-periodic non-radial $g$-mode pulsations 
that --- being global --- can be used to measure
their internal properties and their rate of evolution \citep{Winget08, Fontaine08, Vauclair13, Althaus10, Corsico19}.

\citet{MR76}
found  G~117-B15A to be variable,
and \citet{K82}
found six simultaneous excited periods in its light curve.
The dominant mode has a period of 215~s, a fractional optical amplitude around 22~mma (milli-modulation amplitude, or parts per thousand),
and is stable in amplitude and phase.
The other, smaller pulsation modes, vary in amplitude
from night to night \citep{K95}, either caused by internal instabilities or unresolved components.
Because the DAVs are normal stars except for their
variability \citep{Robinson79, Bergeron95, Bergeron04, Castanheira13, Romero13},
i.e., an evolutionary stage in the cooling of all white dwarfs,
it is likely that the DAV structural properties are representative
of {\it all} hydrogen atmosphere --- DA --- white dwarfs. DA white dwarfs comprise more than 80\% of all white dwarfs
\citep[e.g.][]{kepler19}.

In their review of the properties of pulsating white dwarfs, \citet{Corsico19} list the 250 ZZ Cetis known at the time. Since then, 39 additional ZZ Cetis have been published \citep{Vincent20}.

We report our continuing study
of the star G~117--B15A, 
one of the hottest of the ZZ Ceti stars.
The rate of change of a pulsation period with time for $g$-mode pulsations
in white dwarf stars
is theoretically directly related to its evolutionary timescale
\citep{WHVH},
allowing us to infer the age of a cool white dwarf.
We have been observing the star since 1974 to measure the
rate of period change with time ($\dot P$)
for the largest amplitude periodicity, at 215~s.
Using all the data obtained from 1974 through 2005, \citet{Kepler05} 
estimated the intrinsic rate of period change
\[\dot P_i = \dot P_{\mathrm{observed}} - \dot P_{\mathrm{pm}} =
(3.79 \pm 0.81) \times 10^{-15} \,\rm s/s\]
The quoted uncertainty
was the intrinsic one from the fit only.

\citet{K84} demonstrated that the observed
variations in the light curve of G~117--B15A 
are due to non-radial {\it g}-mode pulsations.
\citet{K00} show the models
predict the effect of radius change due to the still ongoing
contraction are an order of magnitude smaller than the
cooling effect on the rate of period change. 

Concerning the expected
stability of pulsation modes, \citet{Hermes17} used \emph{Kepler} and
\emph{K2} data to show that modes with periods longer than about 800~s
are considerably less coherent than shorter period modes, with their 
power spectra often having a ``mottled'' appearance. 
\citet{mikemon20} showed that this could be explained by the longer period 
modes having a stronger interaction with the surface convection zone of the 
star, since they have turning points much closer to the surface 
than low-period modes. We return to this question with regards to G~117-B15A 
in Section~\ref{phase}.

G~117--B15A is proving to be a useful laboratory for particle
physics \citep{Isern}.
\citet{Corsico2001}
calculated the limit on the axion mass compatible with the
then observed upper limit to the cooling, 
showing $m_a \cos\beta\leq 4.4~\mbox{meV}$ and \citet{K04}
demonstrates axion cooling would be dominant over neutrino
cooling for the lukewarm white dwarf stars for axion masses
of this order.
\citet{Biesiada}
show that the $2\sigma$ upper limit published in \citet{K00}
limits the string mass scale $M_S \geq 14.3~\mbox{TeV}/c^2$ for 6 dimensions,
from the observed cooling rate and the emission of Kaluza-Klein gravitons,
but
the value is unconstrained for higher dimensions.
\citet{Benvenuto04} show the
observed rates of period change can also be used to constrain the
dynamical rate of change of the constant of gravity $\dot G$.

\section{Observations} \label{sec:obs}

\cite{Kepler05} reported on the observations from 1974 to 2005.
In this paper we report on 178~h of
additional time series photometry from 2005 to 2020
(Table~\ref{table:obs}),
most taken with the Argos prime-focus CCD camera  \citep{NA}
on the $2.1$~m Otto Struve telescope at McDonald Observatory.

We use the BG40 filter on all observations because it increases the contrast between the
(mostly blue) pulsational amplitude and the (mostly red) sky
background. Also, the sky background is variable, and introduces
strong systematics.
Non-radial
{\it g}-mode light variations have the same phase in all colors
\citep{RKN} but
the amplitudes decrease
with wavelength. For example, a filter-less observation with
Argos gives an amplitude around 40\% smaller for G~117--B15A.

\begin{table}[h!b] 
\caption{Journal of Observations since 2005}
\label{table:obs}
\hspace*{-4.5em}
\begin{small}
\begin{tabular}{|l|r|r|r|l|} \hline
 Date         &Exposure&  Duration&      Number&     Telescope\cr 
              &(s)     &  (s)     &       -    &        -      \cr
2005-Dec-05  &  5   &     7040    &  1408  & McD 2.1m    \cr 
2005-Dec-09  &  5   &    12000    &  2400  & McD 2.1m    \cr 
2006-Mar-01  &  5   &    14020    &  2804  & McD 2.1m     \cr 
2006-Mar-04  &  5   &     9305    &  1861  & McD 2.1m    \cr 
2006-Mar-06  & 10   &    13140    &  1314  & McD 2.1m     \cr 
2006-Dec-21  & 10   &     4280    &   428  & McD 2.1m     \cr 
2006-Dec-28  &  5   &    15730    &  3147  & McD 2.1m     \cr 
2007-Mar-16  &  5   &     2865    &   545  & McD 2.1m     \cr 
2007-Apr-10  &  5   &     6775    &  1355  & McD 2.1m     \cr 
2008-Feb-09  &  5   &    16040    &  3208  & McD 2.1m     \cr 
2008-Mar-11  &  5   &    10175    &  2035  & McD 2.1m     \cr 
2008-Mar-13  &  5   &    9900     &  1980  & McD 2.1m     \cr 
2008-Mar-14  & 5	&    3600	  &   720  & McD 2.1m     \cr 
2009-Jan-29  &  5   &   15035     &  3008  & McD 2.1m     \cr 
2009-Jan-30  &  5   &   14400     &  2881  & McD 2.1m     \cr 
2009-Feb-03  &  5   &   10315     &  2064  & McD 2.1m     \cr 
2009-Apr-13  &  20  &    7400     &   370  & 0.6m Suhora  \cr 
2009-Apr-15  &  22  &    6974     &   317  & 0.6m Suhora  \cr 
2009-Apr-24  & 10   &   9030      &   903  & McD 2.1m     \cr 
2010-Jan-10 &   5   &   3840      &   768  & McD 2.1m     \cr 
2010-Jan-20 &   5   &    7320     &  1465  & McD 2.1m     \cr 
2010-Feb-16  &  5   &   10805     &  2161  & McD 2.1m     \cr 
2010-Feb-19  & 10   &   10740     &  1075  & McD 0.9m     \cr 
2010-Mar-10  & 15   &   7380      &   493  & McD 0.9m     \cr 
2010-Mar-11  & 15   &   18480     &  1233  & McD 0.9m     \cr 
2010-Mar-17  & 10   &   18350     &  1836  & McD 0.9m     \cr 
2010-Mar-18  & 10   &   27030     &  2704  & McD 0.9m     \cr 
2010-Mar-19  & 10   &   22910     &  2292  & McD 0.9m     \cr 
2010-Mar-21  & 15   &   20160     &  1345  & McD 0.9m     \cr 
2010-Apr-10  & 10   &   14640     &  1464  & McD 2.1m     \cr 
2010-Apr-16  & 10   &    1360     &   137  & McD 2.1m     \cr 
2010-Dec-12  & 10   &   17530     &  1753  & McD 2.1m     \cr 
2011-Jan-06  & 10   &    5810     &   581  & McD 2.1m     \cr 
2011-Jan-09  & 10   &    9770     &   977  & McD 2.1m     \cr 
2011-Feb-01  & 10   &    9870     &   987  & McD 2.1m     \cr 
2013-Apr-13  &  5   &    7900     &  1580  & McD 2.1m     \cr 
2013-Apr-23  &  5   &   10800     &  2160  & McD 2.1m     \cr 
2013-Apr-25  &  5   &   12670     &  2534  & McD 2.1m     \cr 
2015-Jan-17  & 30   &   30240     &   890  & McD 2.1m     \cr 
2015-Jan-18  & 30   &   10620     &   344  & McD 2.1m     \cr 
2015-Mar-14  & 10   &   11070     &  1101  & McD 2.1m     \cr 
2015-Mar-16  & 20   &   18700     &   826  & McD 2.1m     \cr 
2015-Mar-19  & 14   &   14266     &  1019  & McD 2.1m     \cr 
2015-Mar-20  & 25   &    8100     &   294  & McD 2.1m     \cr 
\hline
\end{tabular}
\end{small}
\end{table}

\begin{table}[h!b] 
\caption{Journal of Observations since 2005 - continuation}
\hspace*{-4.5em}
\begin{small}
\begin{tabular}{|l|r|r|r|l|}\hline
 Date         &Exposure&  Duration&      Number&     Telescope\cr 
              &(s)    &  (s)      &       -    &        -      \cr
2018-Jan-26  & 15   &   15720     &  1049  & McD 2.1m     \cr 
2018-Jan-27  & 5    &   14285     &  2856  & McD 2.1m     \cr 
2018-Jan-28  & 10   &   12460     &  1247  & McD 2.1m     \cr 
2018-Mar-12  & 10   &   5380      &   539  & McD 2.1m     \cr 
2018-Mar-13  & 10   &   12480     &   742  & McD 2.1m     \cr 
2018-Mar-14  & 10   &   9710      &   673  & McD 2.1m     \cr 
2018-Mar-15a & 10   &   4350      &   435  & McD 2.1m     \cr 
2018-Mar-15b & 10   &    1940     &   195  & McD 2.1m     \cr 
2019-Jan-05  & 3    &    20142    &  6713  & McD 2.1m     \cr 
2020-Feb-20  & 20   &    8400     &   380  & McD 2.1m     \cr 
2020-Feb-23  & 15   &    18960    &  1132  & McD 2.1m     \cr 
\hline
\end{tabular}
\end{small}
\end{table}
\section{Data Reduction}
We reduce and analyze the data in the manner described by 
\citet{N90},
and \citet{K93}.
We bring all the data to the same fractional amplitude scale,
and the times from terrestrial UTC
to the uniform Barycentric Julian Coordinated Date
(TCB) scale, using JPL DE405 ephemeris \citep{Standish98,Standish04}
to model Earth's motion.
We compute Fourier transforms for each individual run,
and verify that the  main pulsation at 215~s
dominates each data set and has an amplitude stable up to 15\%,
our uncertainty in amplitude due to the lack of accurate
time and color-dependent extinction determination.

\section{Time Scale for Period Change}

As the dominant pulsation mode at $P=215$~s has been stable in frequency and amplitude
since our first observations in 1974, we can calculate the
time of maximum for each new run and look for deviations from those
assuming a constant period.

We fit our observed time of maximum light, $O$, to the equation for the difference to the calculated one, $C$:
\[(O-C) = \Delta E_0 + \Delta P \cdot E
+ \frac{1}{2} P \cdot \dot P \cdot E^2\]
where $\Delta E_0 = (T_{\rm max}^0 - T_{\rm max}^1)$, $\Delta P = (P - P_{t=T_{\rm max}^0})$, and $E$ is the epoch of the time of maximum $T_\mathrm{max}$, i.e, the integer number of cycles
after our first observation $T^0_\mathrm{max}$, which occurred in 16 Dec 1974{\footnote{Fitting the
whole light curve with a term proportional to
$\sin\left[\frac{2\pi}{\left(P+\frac{1}{2}\dot{P}\right)}t + \phi\right]$ by non-linear least squares gives unreliable uncertainty estimates
and the alias space in P and $\dot{P}$ is extremely dense
due to 45\,yr data set span \citep{Darragh}.}}.

\begin{figure}
\begin{center}
	\includegraphics[width=0.35\textwidth,angle=270]{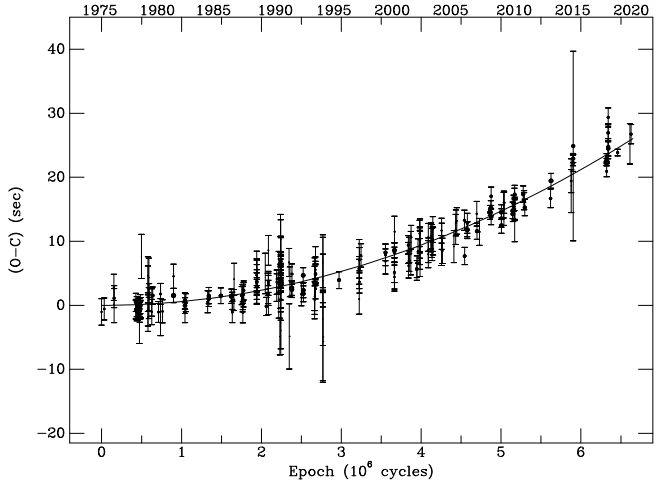}
\caption{
{\bf (O-C)}: {\bf O}bserved minus {\bf C}alculated times of maxima for
the 215~s pulsation of G~117-B15A. 
The size of each point is proportional to its weight, i.e., inversely proportional to the uncertainty in the time of maxima squared. 
We show $\pm 1 \sigma$ error bars for each point,
and the line shows our best fit parabola to the data.
The fact the line does not overlap these error bars are
a demonstration they are underestimate.
Note that as the period of pulsation is 215.1973882~s, the observed total change in phase is
only $50\deg$.}
     \label{o-c}
     \end{center}
\end{figure}

In Figure~\ref{o-c}, we show the O--C timings 
after subtracting the correction to period and epoch,
and our best fit curve through the data.
The size of each point is proportional to its weight,
i.e., inversely proportional to the square of
uncertainty in phase. The error bars plotted are $\pm 1\sigma$.
From our data through 2020, we obtain a new value for the epoch of maximum,
$T_{\rm max}^0 = 244\,2397.9175141 \,{\rm TCB} \pm 0.41 \,\rm s$,
a new value for the period, $P = 215.197 388 23 \pm 0.000 000 63 \,\rm s$,
and most importantly, an observed rate of period change of:
\[\dot P_{\rm obs} = (5.47 \pm 0.82) \times 10^{-15} \,\rm s/s.\]
Our quoted uncertainty is the most conservative estimate from the weighted average,
which accounts for the reduction on the number of effective data points.
For a comparison with the uncertainty published in \citet{Kepler05}, the internal uncertainty is now $0.32\times 10^{-15}$, clearly underestimated from the changes in the value itself.

We use linear least squares to make our fit, with each point weighted inversely
proportional to the uncertainty in the time of maxima for each 
individual run squared.
We quadratically add an additional 1~s of uncertainty to the time of maxima
for
each night to account for external uncertainty caused perhaps
by the beating of possible small amplitude pulsations \citep{K95}
or the small modulation seen in Figure~\ref{o-c}.
The amplitude, 1~s, is chosen as $4\langle A \rangle$ from the Fourier transform of the
$(O-C)$.
Such external uncertainty is consistent with \citet{Splaver} who
show that the true uncertainties of the
times of arrival of the milli-second pulsars are generally larger
than the formal uncertainties, and that a quadratic term is
added to them to fit the observations. 

The satellite {\em TESS} observed G~115-B15A in Sector 21, almost continuously from 21 Jan 2020 to 18 Feb 2020. As the data is co-added on board to 120~s, and the camera is only 15~cm, the observed light curve resulted in an uncertainty of 13.8~s on the time of maximum of the 215~s pulsation. Even though the phase is in agreement with the observed $O-C$, it did not produce any improvement in our \Pdot\ determination.
Including the data from {\em TESS}, the values are unchanged, due to its large uncertainty. We note that TESS data in Sector 21 already includes the correction of 2~s  to Data Product Timestamps in the pipeline\footnote{\url{ https://archive.stsci.edu/missions/tess/doc/tess_drn/tess_reprocessing-sector_14_19_drn30_v01.pdf}}, but there is still an extra uncertainty perhaps as large as 4~s in the {\em TESS} timings, compared to ground observations \citep{VonEssen}.


\section{Discussion}

We claim that the 215\,s periodicity in G~117-B15A is the most stable optical clock known.
According to \citet{Nicholson15}, their
optical atomic clock based on 
2,000 ultracold strontium atoms trapped in a laser lattice
lose no more than 1 second in 15 billion years, with
an accuracy of
$\dot{P}\leq 2 \times 10^{-18}$~s/s in the JILA $^{87}\mathrm{Sr}$ clock.
Considering its period is $2.5\times 10^{-15}$~s, even though it is many orders of magnitude more accurate
than G~117-B15A, it less stable, as its
timescale for period changes, i.e., the time it takes to lose a whole cycle, $P/\dot{P}$ is 1250~s, compared
to 1.2~Gyr for G~117--B15A. The total 26~s phase change observed for G~117-B15A after 45 years of observations implies one cycle of the phase will be reached in 372.5~yr.
In terms of accuracy, \citet{Brewer19} reports the NIST $^{27}\mathrm{Al}^+$ quantum-logic clock reached a systematic uncertainty of $\dot{P}\simeq 9.4\times 10^{-19}$.
Even the Hulse \& Taylor's millisecond pulsar
\citep{Hulse}, has a timescale for period change $P/\dot{P}$ of only 0.35~Gyr
\citep{Damour}, but the radio millisecond pulsar
PSR~J1909-3744 \citep{Liu20} has $\dot{P}_i=2.60(3) \times 10^{-21}$~s/s,
and a timescale of 18~Gyr, after 15 years of observations.
after correcting for the motion effects (pulsar proper motion, galactic differential acceleration, orbital motion and general relativity correction), jitter, red and white noise models.
The timescale based on the spin of radio pulsars with millisecond periods can have a stability comparable to that of atomic timescales, but millisecond pulsars are also known to undergo sudden small glitch events \citep[e.g.][]{McKee16}, magnetospheric changes \citep{Shannon16} and effects relating to sudden changes in the interstellar medium \citep{Lentati16, Brook18, Lam18}. 

G~117--B15A was the first pulsating white dwarf to have its main
pulsation mode index identified. The 215~s mode has $\ell=1$, 
as determined by comparing the
ultraviolet pulsation amplitude, measured with the Hubble
Space Telescope, to the optical amplitude \citep{R95}.
Using time-resolved spectra obtained at the Keck Telescope,
\citet{Kotak} confirm the $\ell$
measurement for the P=215~s pulsation and show that the
other large amplitude modes, at 271~s and 304~s,
show chromatic amplitude changes that do not fit simple single mode
theoretical models \citep{R95}. 
\citet{R95}, and \citet{KAV}
derive $T_{\mathrm{eff}}$ near 12,400~K, 
while \citet{Bergeron95,Bergeron04}
using a less efficient model for convection, 
estimated $T_{\mathrm{eff}}$=11,630~K.
\citet{Gianninas11} used ML2/$\alpha=0.8$ models, which corrected to
\citet{Tremblay13} tri-dimensional convection calculations
correspond to $T_\mathrm{eff}=12\,420$~K, and $\log g=8.12$. 
The uncertainty in
effective temperature determinations from spectroscopy are of the order of 300~K,
and 0.05~dex in the surface gravity \citep{Bergeron95}.

\citet{Benvenuto02} show the seismological models with time-dependent
element diffusion are only consistent
with the spectroscopic data if the modes are the $\ell=1$, $k=2$, 3, and
4,
and deduces $M=0.525~M_\odot$, $\log(M_H/M_\star)\geq -3.83$ and
$T_{\mathrm{eff}}=11\,800$~K, similar to those by \citet{KA00}.
Their best model predicted a parallax $\Pi$=15.89~mas,
$\dot{P}=4.43 \times 10^{-15}$~s/s, for
the P=215~s,
$\dot{P}=3.22 \times 10^{-15}$~s/s, for
the P=271~s, and
$\dot{P}=5.76 \times 10^{-15}$~s/s, for
the P=304~s periodicities.

\citet{Romero12} used the mode identification and the 
observed periods of the three largest known
pulsation modes to solve earlier degeneracy in solutions and derive a hydrogen layer mass 
best estimate of $1.25 \times 10^{-6}\,M_\star$,
assuming $k=2$ for the 215~s mode on their evolutionary C/O core white dwarfs,
which resulted in C/O=0.28/0.70 for its mass. The
core composition is constrained mainly by
the presence of the 304~s pulsation.
In their Table~2, \citet{Corsico12a} quote the theoretical rates of period change for the \citet{Romero12} best fit model as $\dot{P}=1.25\times 10^{-15}$, $4.43\times 10^{-15}$, and $4.31\times 10^{-15}$, for the $k=2$, 3 and 4 modes.
The $k=2$ mode corresponds to the $P=215$~s trapped mode in the hydrogen layer. Similar values were found by \citet{Kim08} for their thicker hydrogen layer solution while their thinner solution had 
$\dot{P} \sim 3.0 \times 10^{-15}$.
\citet{Corsico12a} also show that because the $k=2$ mode is trapped
at the surface hydrogen layer, its rate of period change is almost insensitive
to the core composition.

While it is true that the period change timescale can be proportional to
the cooling timescale, it is also possible that other phenomena with
shorter timescales can affect $\dot P$. The cooling timescale is the longest
possible one.

As a corollary, if the observed $\dot P$ is low enough
to be consistent with evolution, then other processes, such as perhaps a
magnetic field or
diffusion induced changes in the boundary layers,
are not present at a level sufficient to affect $\dot P$.


\subsection{Theoretical Estimates and Corrections}

\subsubsection{Proper Motion}
Stars are moving --- they are observed to have a proper motion across the sky. As shown by 
\citet{Shklovskii70}, and known as the ``Shklovskii effect'', this means that 
the observed period derivatives will be higher than the intrinsic period derivative
by an amount proportional to $v^2/cd$.
\citet{Pajdosz95} estimated the influence of the proper motion of the star
on the measured $\dot P$ as:
\[\dot P_{\mathrm{obs}} = \dot P_{\mathrm{evol}}\left(1+v_r/c\right)
+ P\dot v_r/c\]
where $v_r$ is the radial velocity of the star. Assuming
$v_r/c \ll 1$
he derived
\[\dot P_{\mathrm{pm}} = 2.430 \times 10^{-18} P[s]
\left(\mu[\,"/yr]\right)^2 d[{\mathrm{pc}}]\]
where $\dot P_{\mathrm{pm}}$ is the effect of the proper motion
on the rate of period change, $P$ is the pulsation period,
$\mu$ is the proper motion and $d$ is the distance.
The proper motion, $\mu=0.1453 \pm 0.0001\,"/{\mathrm{yr}}$,
and the parallax,
$\Pi=(0.01739 \pm 0.0008)\,"$,
were estimated by Gaia DR2, for both G~117-B15A and its proper motion companion G~117-B15B:
\[A: \pi=17.386\pm 0.080\,\mathrm{mas} \quad  d=57.5\pm 0.2\,\mathrm{pc}\]
\[\mu=(-145.30\pm 0.10,-0.006\pm 0.088)\,\mathrm{mas/yr}\]
\[B: \pi=17.437\pm 0.101\,\mathrm{mas}\]\[ \quad \mu=(-145.99\pm 0.12,-0.290\pm 0.112)\,\mathrm{mas/yr}\]
Therefore
$\dot P_{\mathrm{pm}}= (0.3532 \pm 0.00024) \times 10^{-15}$~s/s,
and the evolutionary ---  intrinsic --- rate of period change $\dot{P}_i$:
\[\dot P_i = \dot P_{\mathrm{observed}} - \dot P_{\mathrm{pm}} = 
(5.12 \pm 0.82) \times 10^{-15} \,\rm s/s\]

\subsubsection{Limits on Mode Coherence}
\label{phase}
\citet{mikemon20} showed that the result of \citet{Hermes17} 
that modes with periods longer than about 800~s
are considerably less coherent than shorter period modes could be explained 
by their interaction with the time-dependent convection zone. Since the modes are assumed
to acquire a small phase shift each time they reflect off the base of the convection
zone, we can estimate the average amount of phase that would be accumulated by the 215~s mode 
over the total time base of observations. While the details are presented in Appendix~\ref{drift},
we find that average accumulated phase would be only $\sim 4 \times 10^{-3}$~rad, which translates into a shift in the $O - C$ diagram of only $\sim 0.13$~s, i..e., negligible.

\subsubsection{Effect of a Changing Magnetic Field}
\label{magfield}

A weak magnetic field can perturb the oscillation frequencies of a star in much the same way that slow rotation does. If this magnetic field also slowly changes its magnitude with time, then it will produce a non-evolutionary \Pdot\ for the modes. Here we provide an estimate of the size and rate of change of the magnetic field that would be required to mimic the observed \Pdot\ for the 215~s mode; details are given in  Appendix~\ref{magpert}.

Employing the same approach as \citet{Jones89} and \citet{Montgomery94}, we find that a uniform magnetic field that decreases from 280~G to 0~G over a time period of 46 years can produce $\Pdot \approx 5.1 \times 10^{-15}$~s/s for a 209~s, $k=2$ mode. In addition, it is the change in $B^2$ that matters, so the same effect would be produced by a field that decreases from 2814~G to 2800~G over a period of 46 years.

\subsection{Pulsation Models\label{models}}
With time, as the temperature in the core
of a white dwarf decreases, 
electron degeneracy increases and the pulsational spectrum of the star shifts to longer periods, in the absence of significant residual gravitational contraction.
We compare the measured value of $\dot{P}_i$ with the range of theoretical values derived from models with C/O cores subject to {\it g}--mode pulsations in the temperature range of G~117--B15A which allow for mode trapping.
\citet{Kim08} estimated for their best model with 
$T_\mathrm{eff} = 12656$~ K, $M_\star = 0.602\,M_\odot$, and a helium layer mass of 
$ 3.55 \times 10^{-3} \, M_\star $
$\dot{P}= (1.92\pm 0.26) \times 10^{-15} \,\rm s/s$ if $\log(M_\mathrm{H}/M_\star) = -6.2$ and
$\dot{P}= (2.98\pm 0,17) \times 10^{-15} \,\rm s/s$ if $\log(M_\mathrm{H}/M_\star) = -7.4$.
The adiabatic pulsation calculations of \citet{Romero12} with realistic evolutionary models, 
give a mass of $0.593~M_\odot$, $\log(M_\mathrm{H}/M_\star)-5.9$ and
$\dot{P}\simeq 1.25 \times 10^{-15} \,\rm s/s$
for the $\ell=1$, $k=2$ observed oscillation.

The observed $P/\dot P =1.33 \times 10^9$ yr
is equivalent to 1~s change in period
in 6.2 million years. We have therefore
measured a rate consistent with the evolutionary time scale
for this lukewarm white dwarf.

\subsubsection{Core Composition}
For a given mass and internal temperature distribution,
theoretical models show that the rate of period change
increases if the mean atomic weight of the core is increased,
for models which have not yet crystallized in their interiors.
As the evolutionary model cools, 
its core crystallizes due to Coulomb
interactions between the ions
\citep{Lamb}, and crystallization  slows down
the cooling by the release of latent heat.
\citet{Mike99} describe the effect of 
crystallization on the pulsations of white dwarf stars,
but G~117--B15A is not cool or massive enough to have a crystallized core
\citep{W97}, or even for the convective coupling of the core to the envelope
described by \citet{Fontaine01} to occur.

The heavier the particles
that compose the  nucleus of the white dwarf, the faster it
cools.
The best estimate of mean atomic weight $A$ of the core comes from the
comparison of the observed $\dot P$ with values from an
evolutionary sequence of white dwarf models. 
\citet{Brassard92} computed the rates of period changes for
800 evolutionary models with various masses,
all with carbon cores but differing He/H surface layer masses, 
obtaining values similar to those
of \citet{W81}, 
\citet{Wood88},
and \citet{Bradley91}.
In those models, the average value of $\dot P$ for all
$\ell=1$, 2 and 3 modes with periods around 215~s in models with an
effective temperature around 13,000~K, and a mass of 0.5~\Msun, is:
$\dot P(\mbox{C core}) =  (4.3 \pm 0.5) \times 10^{-15}\,\rm s/s.$
\citet{Benvenuto04} C/O models give 
$\dot P(\mbox{C/O core}) =  (3-4) \times 10^{-15}\,\rm s/s.$
Using a Mestel-like cooling law \citep{Mestel52, Kawaler86}, i.e.,
$\dot T \propto A$, where $A$ is the mean atomic weight in the core,
one could write, for untrapped modes:
\[\dot P(A) =  (3-4) \times 10^{-15}\,\frac{A}{14} \,\rm s/s.\]
All these models were computed assuming a thick $\log(M_H/M_\star)=10^{-4}$ hydrogen layer, which lead to no significant mode trapping.
The observed rate of period change is
therefore consistent with a C or C/O core. The largest uncertainty comes
from the models, essentially the hydrogen layer mass \citep{Kim08}.

\subsubsection{Reflex Motion}
The presence of an orbital companion
could contribute to the period change we have detected.
When a star has an orbital companion, the variation of its line-of-sight
position with time produces a variation in the time of arrival of
the pulsation maxima, by changing the light travel time between the star
and the observer by reflex motion of
the white dwarf around the barycenter of the system.
\citet{K91} estimated a contribution to $\dot{P}$ caused by
reflex orbital motion of 
the observed proper motion companion of G~117--B15A in
their equation (10) as:
\[\dot{P}_{\mathrm{orbital}} = \frac{P_{\mathrm{pul}}}{c} 
\frac{GM_B}{a_T^2}
= 1.97 \times 10^{-11} P_{\mathrm{pul}}\,
\frac{M_B/M_\odot}{(a_T/AU)^2}\,{\mathrm{s/s}}\]
where
$a_T$ is 
the total separation, $G$ here is the gravitational constant, $M_B$ is the mass of the companion star.
In the above derivation
they have also assumed
the orbit to be nearly edge on to give the largest effect possible.
G~117-B15A with Gaia magnitude $G=15.5589\pm 0.0010$, absolute magnitude $M_G=11.760$, and Gaia color $G_{\rm BP}-G_{\rm RP}=-0.020$,
Gaia DR2 parallax $\pi=(17.39\pm 0.08)$~mas, proper motion ppm=$(145.34\pm 0.10,-0.01\pm 0.09)$~mas/yr,
and its common proper motion companion
\object[G 117-B15B]{G~117--B15B},
with $G=14.7270\pm 0.0010$, $M_G$=10.934, BP-RP=2.885, 13.8" away,
$\pi=17.43\pm 0.10$~mas, ppm=$(-145.99 0.12, -0.29 0.11)$~mas/yr,
are a common proper motion pair, 
forming a real binary system.
\citet{Silvestri}
measured the radial velocity of G~117-B15B, assuming it formed
a wide binary system with G~117-B15A as only $v_r=2.2\pm 9.4$~km/s.
\citet{Kotak} classifies G~117-B15B as an M3Ve from its spectra,
obtained with the 10~m Keck I telescope,
and measured $\log g\simeq 4.5$ and $T_{\mathrm{eff}}\simeq 3400$~K.
\citet{Kirkpatrick11} classifies G117-B15B as M3.5V from WISE colors.
The mass of an M3.5V should be around $0.33~M_\odot$ \citep{Lang}.\ 
With a separation of 13.8~arcsec, 
$a_T= 794$~AU, assuming the observed distance between G~117-B15A and B is at its largest ($\sin \omega\simeq 1$),
where $\omega$ is the argument of periapsis. This corresponds to a lower limit on the orbital period of around 22\,000 years,
and we estimate 
$\dot P_{\mathrm{orbital}} \leq (1.1 \pm 1.1) \times 10^{-17}\,{\mathrm{s/s}}$.
The large uncertainty takes into account the possibility the orbit might
be strongly elliptical.
Even though G~117--B15A and B form a real binary system, the contribution of the orbital 
reflex motion to the observed \Pdot\ is negligible.

The whole observed phase change 
{\it could} also be caused
by a planet of Jupiter's mass orbiting the white dwarf
edge-on at a distance of 31~AU, which
corresponds to an orbital period around 314~yr, 
or a more massive planet in a less inclined orbit.
\citet{Duncan} show that such a planet would survive the post-main
sequence mass loss.
Any closer to the white dwarf, and such planets would produce 
a larger $\dot{P}$ \citep[e.g.][]{Krzesinski20}.
Note however that reflex motion produces {\it sinusoidal} variations on
the $O-C$, which are distinguishable from parabolic variations
after a significant portion
of the orbit has been covered. 
This allows us to rule out the presence of planets as a function of
orbital period and $M\sin{i}$, where $i$ is the orbital inclination
\citep[see Figure~4 in][]{Mullally08}.
Considering a second-order derivative of the $(O-C)$
has not been detected yet, only planets with orbital periods longer than about 
900~yr should be indistinguishable from a parabola, or if their effect on the (O-C) is smaller than 1~s, i.e.,
with $M\sin{i}$ similar to the Earth's mass. 
The theoretical upper limit for a stable planetary orbit around G~117-B15A is 
around $0.3~a_T$ \citep{Musielak05}, i.e., around 240~AU, assuming the 
observed distance between G~117-B15A and B, $a_T$, is at its largest ($\sin \omega\simeq 1$),
which would lead to a period of 4800~yr. 
At that distance, a planet would have to be more massive than $2.3~M_J$ 
to produce a phase change in 45~yr as large as the 26~s observed. Note that for half of the
orbit the correction has the opposite sign. If the $\dot{P}$ measured for other ZZ Cetis,
like R548 \citep{Mukadam13} and L19-2 \citep{Sullivan15} are also larger than the white dwarf cooling timescales, it is unlikely they are all caused by planets
traveling away from us.

As discussed by \citet{Damour}, any relative acceleration of the
star with respect to the barycenter of the solar system will contribute
to the observed $\dot{P}$. Their equations (2.2) for the
differential galactic orbits, decomposed in
a planar contribution (2.12),
where the second term is the proper motion correction, 
and a perpendicular contribution (2.28),
applied to G~117-B15A, show the galactic contribution to be exactly the
one calculated above for proper motion, i.e., the other terms are
negligible --- 2 to 3 orders of magnitude smaller.

\subsection{Axions}

In section~\ref{models} we list the predicted value of $\dot{P}$ for the $k=2, \ell=1$ trapped mode for the evolutionary models as
$\dot{P}\simeq 1.25\times 10^{-15}$. 
As the value of the observed rate of period change is larger than the theoretical model, we study the possibility of the excess of cooling as due to axions --- hypothetical weakly interacting particles proposed as a solution
to the strong charge-parity problem in quantum chromodynamics \citep{PQ}.
This possibility was first raised by \citet{Isern92} since axions, similar to neutrinos, can escape carrying energy. At the time, employing semi analytical models to the observed period change of G~117-B15A ($\dot{P}=12\pm 3.5\times10^{-15}$), they estimated a mass of $m_a \simeq 8.7$~meV. \citet{K00} published a value for $\dot{P}$ ($2.3\pm 1.4\times 10^{-15}$), much lower than the previous value, and \citet{Corsico2001} estimated $m_a<4.4$~meV using a detailed asteroseismological model. Later, with improved models and determination of $\dot{P}$ \citep{Kepler05}, \citet{Kim08} estimated $m_a < 13.5$~meV. The determination of $\dot{P}$ by \citet{Kepler12} was used by \citet{Corsico12a} to estimate $m_a\simeq 17.4$~meV. This idea was also applied to other DAVs with $\dot{P}$ known, finding $m_a\simeq 17.1$~meV for R~548 \citep{Corsico12b} and $m_a<25$~meV for L~19-2 \citep{Corsico16}.


Now, using the new determination for $\dot{P}$ of G117-B15A we are able to set new constraints on the axion mass, assuming the extra cooling is due to the putative axion.


\begin{figure}
\begin{center}
	\includegraphics[width=0.47\textwidth]{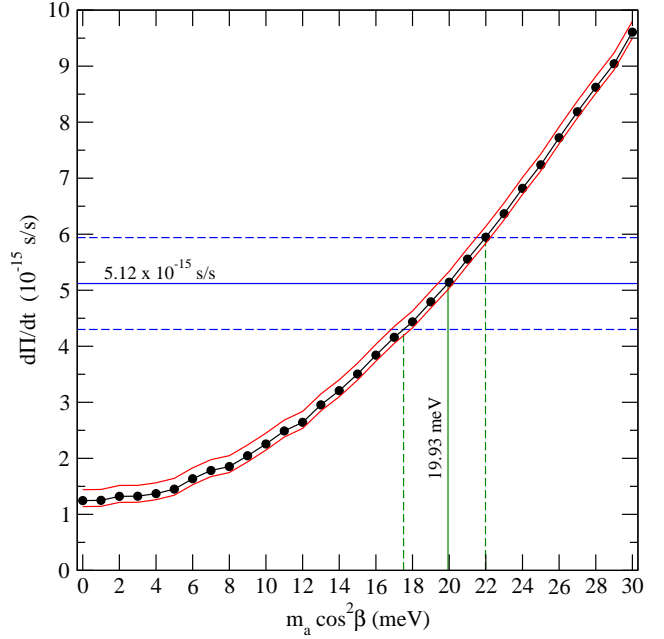}
     \caption{The rate of period change for the mode with $\ell = 1$ and $k=2$, corresponding to a period of $\sim 215$~s in term of the axion mass (black circles). Dashed lines represent the uncertainties in the value in the observed  $\dot{P}$ and the axion mass, while the red curves represent the internal uncertainties in $\dot{P}$ due to modeling.   }
     \label{classes}
     \end{center}
\end{figure}
Using the value for the intrinsic $\dot{P}$, and assuming the effects from possible orbiting planets
and magnetic fields are negligible, we estimate an axion mass using fully evolutionary models (see Fig.~\ref{classes}) calculated with {LPCODE}
\citep[e.g.][]{Althaus10, Romero12} including axions during all the white dwarf cooling. These results are very similar to those of Fig.~5 of \citet{Corsico12b}.
As a result we obtain a value of the coupling constant between axions and electrons $g_{ae}=(5.66\pm 0.57) \times10^{-13}$, or, adopting the DFSZ model \citep{DFZ, DFSZ}, an axion mass $m_a \cos^2 \beta = 20\pm 2$~meV.
\citet{deGeronimo17}  estimated the high and low limits for the C($\alpha,\gamma$) O reaction rate from the uncertainties given in \citet{Kunz02}.  These limits are $0.55\times$ and $1.1\times$ factors in the reaction rate, which translates into a central carbon abundance of $X_C = 0.450$ and $X_C = 0.246$ respectively, for the best fit model for G~117-B15A. For the k=2, l=1 mode, corresponding to the 215~s mode, the value for $\dot{P}$, in the case where no axions are considered, changes by $\sim 15$\% and $\sim 9$\%, respectively. Since this mode is trapped in the envelope, we do not expect large differences in the value of the rate of period change  when the central composition changes \citep{Corsico16}. Considering both the observational and model uncertainties, the estimated axion mass is $m_a \cos^2 \beta =19.9_{-3.1}^{+2.1}$~meV.

\section{Conclusions}
We have measured the rate of change of the main pulsation period
for the $T_{\mathrm{eff}} \simeq 12\,400$~K pulsating DA white dwarf
G~117--B15A, the first ZZ Ceti to have its evolutionary rate of
change measured, confirming it is the most stable optical clock
known, with a rate of change of 1~s in $\simeq 6.2$ million years and
a precise laboratory for physics at high energy.
We note that mode trapping can reduce
the rate of period change by up to a factor of two \citep{Bradley96,Corsico12a},
but the changes in the trapping layers are still caused
by cooling, and are included in our theoretical models.

After a large investment of telescope time to achieve such
precision, we have measured the cooling rate of this
 2.16 Gyr old white dwarf \citep{Romero12} --- or 1.79~Gyr for our models with
 20~meV axions. This estimate includes the time the star, with 
 $M_{\rm initial}\simeq 1.75\,M_\odot$ \citep{Romero12},
took to reach the white dwarf phase. We have also demonstrated
it does not harbor planetary bodies similar to Jupiter in mass
up to a distance around 30~AU from the star, modulo $\sin i$, where $i$ is the inclination
to the line of sight. We cannot exclude larger distances or smaller planets
with light travel time effects on the white dwarf smaller than 1s.

%

\acknowledgments
This work was partially supported by
grants from CNPq (Brazil), CAPES (Brazil), FAPERGS (Brazil)
NSF (USA),  NASA (USA).
Z.P.V., D.E.W., and M.H.M. acknowledge support from the United States Department of Energy under grant DE-SC0010623, the National Science Foundation under grant AST 1707419, and the Wootton Center for Astrophysical Plasma Properties under the United States Department of Energy collaborative agreement DE-NA0003843. M.H.M. acknowledges support from the NASA ADAP program under grant 80NSSC20K0455. We acknowledge support through the {\em TESS} Guest Investigator Program Grant 80NSSC19K0378. K.J.B.\ is supported by the National Science Foundation under Award AST-1903828.
This work has made use
of data from the European Space Agency (ESA) mission Gaia (https://www.cosmos.esa.int/gaia), processed by the Gaia Data Processing and Analysis Consortium (DPAC, https://www.cosmos.esa.int/web/gaia/dpac/consortium). Funding for the DPAC has
been provided by national institutions, in particular the
institutions participating in the Gaia Multilateral Agreement. 
This paper includes data collected by the {\em TESS} mission. Funding for the {\em TESS} mission is provided by the NASA Explorer Program.
We made extensive use of NASA Astrophysics
Data System Bibliographic Service (ADS) and the SIMBAD and VizieR database, operated at CDS, Strasbourg,
France.


\begin{table}
\begin{center}
\caption{Total Data Set to Date}
\begin{tabular}{|r|l|c|c|}  \hline
{\bf Time of Maximum} & {\bf Epoch of}   & \boldmath$(O-C)$  & \boldmath$\sigma$   \\
BJDD                  & {\bf Maximum}    &(sec)              & (sec)                \\ \hline
2442397.917507 &       0 &  0.0 & 2.1 \\ 
2442477.797089 &   32071 &  0.5 & 1.7 \\ 
2442779.887934 &  153358 &  3.9 & 2.1 \\ 
2442783.850624 &  154949 &  1.2 & 2.9 \\ 
2442786.981458 &  156206 &  2.2 & 1.5 \\ 
2443462.962774 &  427607 &  1.6 & 1.4 \\ 
2443463.946592 &  428002 &  0.5 & 1.4 \\ 
2443465.969049 &  428814 &  0.5 & 1.6 \\ 
2443489.909755 &  438426 &  0.2 & 1.5 \\ 
2443492.898616 &  439626 &  0.9 & 1.6 \\ 
2443521.927837 &  451281 &  0.1 & 1.3 \\ 
2443552.752879 &  463657 &  0.8 & 1.4 \\ 
2443576.725940 &  473282 & -1.6 & 3.3 \\ 
2443581.692438 &  475276 &  0.3 & 1.3 \\ 
2443582.693698 &  475678 & -0.2 & 1.3 \\ 
2443583.697469 &  476081 &  1.0 & 1.3 \\ 
2443584.733602 &  476497 &  0.8 & 1.4 \\ 
2443604.659292 &  484497 &  1.3 & 1.5 \\ 
2443605.752703 &  484936 &  0.4 & 1.4 \\ 
2443611.693050 &  487321 &  0.6 & 1.3 \\ 
2443613.658222 &  488110 &  0.7 & 1.6 \\ 
2443636.674971 &  497351 &  8.8 & 3.4 \\ 
2443839.956765 &  578967 &  5.8 & 3.0 \\ 
2443841.976708 &  579778 &  3.7 & 3.5 \\ 
2443842.980413 &  580181 & -0.7 & 2.2 \\ 
2443843.944332 &  580568 &  0.5 & 2.6 \\ 
2443869.989703 &  591025 &  1.5 & 2.4 \\ 
2443870.946182 &  591409 &  5.5 & 3.1 \\ 
2443874.916339 &  593003 &  2.4 & 2.1 \\ 
2443959.695117 &  627041 &  0.1 & 2.0 \\ 
2443963.662836 &  628634 &  1.6 & 2.1 \\ 
2443990.664641 &  639475 &  2.7 & 1.3 \\ 
2444169.945954 &  711455 &  0.1 & 1.6 \\ 
2444231.822666 &  736298 & -0.7 & 2.9 \\ 
2444232.818992 &  736698 &  3.0 & 1.6 \\ 
2444293.833896 &  761195 &  0.3 & 1.8 \\ 
2444637.776174 &  899285 &  5.8 & 1.9 \\ 
2444641.624287 &  900830 &  2.8 & 1.1 \\ 
2444992.789531 & 1041820 &  0.1 & 1.6 \\ 
2444994.689956 & 1042583 &  1.2 & 1.2 \\ 
2444996.744801 & 1043408 &  2.0 & 1.3 \\ 
2444997.723649 & 1043801 &  1.9 & 1.2 \\
2445021.716661 & 1053434 &  1.7 & 1.4 \\ 
2445703.860004 & 1327309 &  1.9 & 1.7 \\ 
2445734.642701 & 1339668 &  2.4 & 1.2 \\ \hline
\end{tabular}
\end{center}
\end{table}

\begin{table}
\begin{center}
\begin{tabular}{|r|l|c|c|}  \hline
{\bf Time of Maximum} & {\bf Epoch}   & {\boldmath$(O-C)$}  & {\boldmath$\sigma$}   \\
BJDD                 &    &(sec)       & (sec)  \\ \hline
2445735.643972 & 1340070 &  2.8 & 1.3 \\
2446113.763716 & 1491882 &  2.9 & 1.2 \\ 
2446443.775386 & 1624379 &  2.8 & 1.1 \\ 
2446468.630178 & 1634358 &  2.1 & 1.3 \\ 
2446473.718679 & 1636401 &  0.3 & 1.6 \\ 
2446523.620086 & 1656436 &  2.2 & 1.6 \\ 
2446524.613917 & 1656835 &  5.5 & 2.5 \\ 
2446768.855451 & 1754896 &  2.9 & 1.4 \\ 
2446794.935676 & 1765367 &  2.5 & 2.1 \\ 
2446796.928219 & 1766167 &  0.3 & 1.6 \\ 
2446797.924535 & 1766567 &  3.1 & 1.3 \\ 
2446798.903378 & 1766960 &  2.6 & 1.8 \\ 
2446823.663537 & 1776901 &  3.1 & 1.9 \\ 
2446825.651132 & 1777699 &  3.7 & 1.5 \\ 
2447231.328096 & 1940575 &  3.7 & 1.9 \\ 
2447231.612054 & 1940689 &  5.1 & 3.5 \\ 
2447232.396626 & 1941004 &  5.0 & 1.6 \\ 
2447232.623291 & 1941095 &  5.9 & 2.9 \\ 
2447233.343090 & 1941384 &  4.5 & 1.3 \\ 
2447233.634506 & 1941501 &  4.7 & 2.3 \\ 
2447234.319475 & 1941776 &  6.8 & 3.2 \\ 
2447235.313250 & 1942175 &  5.2 & 1.4 \\ 
2447235.607168 & 1942293 &  6.4 & 2.1 \\ 
2447236.610922 & 1942696 &  6.2 & 1.6 \\ 
2447589.375198 & 2084328 &  3.2 & 1.4 \\ 
2447594.331735 & 2086318 &  5.2 & 1.6 \\ 
2447595.323018 & 2086716 &  3.5 & 2.0 \\ 
2447596.311907 & 2087113 & 10.1 & 2.3 \\ 
2447597.315602 & 2087516 &  4.8 & 1.7 \\ 
2447598.319339 & 2087919 &  3.1 & 3.1 \\ 
2447499.072036 & 2048072 &  6.5 & 3.2 \\ 
2447532.768799 & 2061601 &  1.3 & 1.4 \\ 
2447853.846325 & 2190511 &  4.3 & 2.1 \\ 
2447856.832697 & 2191710 &  5.2 & 1.9 \\ 
2447918.644630 & 2216527 &  2.6 & 3.1 \\ 
2447920.619811 & 2217320 &  6.7 & 3.3 \\ 
2447952.622834 & 2230169 & -3.3 & 2.9 \\ 
2447972.620899 & 2238198 &  9.6 & 6.1 \\ 
2447973.709340 & 2238635 &  9.7 & 2.6 \\
2447973.741682 & 2238648 &  6.5 & 1.4 \\ 
2447978.770467 & 2240667 & 10.0 & 2.1 \\ 
2447979.781717 & 2241073 & 11.8 & 3.1 \\ 
2447980.319627 & 2241289 &  4.6 & 3.5 \\ 
2447977.403038 & 2240118 &  7.5 & 2.3 \\
2447978.327055 & 2240489 &  4.3 & 3.3 \\ 
2447979.358189 & 2240903 &  2.6 & 3.4 \\ 
2447979.358145 & 2240903 & -1.2 & 4.9 \\ 
2447978.601069 & 2240599 &  7.4 & 2.5 \\ \hline
\end{tabular}
\end{center}
\end{table}

\begin{table}
\begin{center}
\begin{tabular}{|r|l|c|c|}  \hline
{\bf Time of Maximum} & {\bf Epoch}   & {\boldmath$(O-C)$}  & {\boldmath$\sigma$}   \\
BJDD                 &    &(sec)       & (sec)  \\ \hline
2447980.621017 & 2241410 &  5.8 & 3.4 \\ 
2447980.782929 & 2241475 &  7.2 & 2.3 \\ 
2447981.325918 & 2241693 &  8.4 & 1.4 \\ 
2447981.592393 & 2241800 &  5.7 & 1.4 \\ 
2447981.779185 & 2241875 &  4.8 & 1.1 \\ 
2447982.329663 & 2242096 &  7.4 & 1.8 \\ 
2447982.743093 & 2242262 &  5.0 & 1.2 \\ 
2447983.734400 & 2242660 &  5.4 & 1.2 \\ 
2447979.281057 & 2240872 &  9.5 & 2.9 \\ 
2447980.224899 & 2241251 & -2.4 & 2.9 \\ 
2447984.735678 & 2243062 &  6.5 & 1.1 \\ 
2448245.724666 & 2347847 & -3.3 & 5.1 \\ 
2448267.799932 & 2356710 &  5.2 & 2.3 \\ 
2448324.627972 & 2379526 &  4.3 & 1.2 \\ 
2448325.708938 & 2379960 &  4.1 & 1.3 \\ 
2448328.593208 & 2381118 &  6.4 & 1.6 \\ 
2448331.661735 & 2382350 &  4.0 & 1.2 \\ 
2448238.571479 & 2344975 &  8.3 & 2.2 \\ 
2448622.833258 & 2499253 &  3.3 & 1.8 \\ 
2448680.642683 & 2522463 &  6.3 & 1.2 \\ 
2448687.614155 & 2525262 &  4.0 & 1.2 \\ 
2448688.597979 & 2525657 &  3.4 & 1.2 \\ 
2449062.660365 & 2675840 &  4.2 & 1.6 \\
2449063.609354 & 2676221 &  6.7 & 1.9 \\ 
2449066.615640 & 2677428 &  6.5 & 1.4 \\ 
2449066.371558 & 2677330 &  7.2 & 2.0 \\ 
2449066.326737 & 2677312 &  8.2 & 2.6 \\ 
2449069.342967 & 2678523 &  6.4 & 1.7 \\ 
2449298.239287 & 2770423 &  8.5 & 4.1 \\ 
2449298.304041 & 2770449 &  8.2 & 4.1 \\ 
2449294.214264 & 2768807 &  5.5 & 4.1 \\ 
2449294.293897 & 2768839 & -0.5 & 4.1 \\ 
2449295.439583 & 2769299 & -4.0 & 6.1 \\ 
2449295.494387 & 2769321 & -3.3 & 7.1 \\ 
2449036.809260 & 2665461 &  2.4 & 2.2 \\ 
2449038.677300 & 2666211 &  3.1 & 2.2 \\ 
2449040.687310 & 2667018 &  3.6 & 4.1 \\ 
2449041.616360 & 2667391 &  4.9 & 4.1 \\ 
2449799.723888 & 2971765 &  5.6 & 1.3 \\ 
2450427.920960 & 3223981 &  8.2 & 3.8 \\ 
2450429.973242 & 3224805 &  2.7 & 2.4 \\ 
2450430.914779 & 3225183 &  6.9 & 2.5 \\ 
2450431.843821 & 3225556 &  7.5 & 1.5 \\ 
2450434.912392 & 3226788 &  8.8 & 2.0 \\ 
2450436.929828 & 3227598 &  5.4 & 1.7 \\ 
2450483.633189 & 3246349 &  9.6 & 1.8 \\
2451249.5989069 & 3553878&10.1&1.3 \\
2451249.7632895 & 3553944&9.7 &1.7 \\ 
\hline
\end{tabular}
\end{center}
\end{table}

\begin{table}
\begin{center}
\begin{tabular}{|r|l|c|c|}  \hline
{\bf Time of Maximum} & {\bf Epoch}   & {\boldmath$(O-C)$}  & {\boldmath$\sigma$}   \\
BJDD                 &    &(sec)       & (sec)  \\ \hline
2451250.6126098 & 3554285&8.7&2.0 \\
2451526.8772586 &3665203& 10.4& 1.2\\
2451528.8523866 &3665996&10.0&1.5\\
2451528.9196061& 3666023&7.4&1.4\\
2451528.9868422& 3666050&6.3&1.9\\
2451529.8585943& 3666400& 6.6&2.0\\
2451530.9097492 & 3666822&    13.4&2.41\\
2451960.8561629 & 3839442&    10.1&1.62\\
2451962.7864775 & 3840217&    11.3&1.48\\
2451967.6806926 & 3842182&    8.1 &1.93\\
2451988.7919772 & 3850658&   10.5 &2.00\\
2451990.7845255 & 3851458&    8.8 &1.59\\
2452037.6472583 & 3870273&   10.1 &3.39\\
2452045.6399770 & 3873482&    12.5&1.85\\
2452225.9050927 & 3945857&    7.6 &1.34\\
2452225.9598927 & 3945879&    8.0 &0.65\\
2452263.8834810 & 3961105&    10.6&0.58\\
2452316.6442205 & 3980721&    13.1&1.0\\
2452317.8995164 & 3982288&    12.2&0.67\\
2452319.7999417 & 3982691&    12.0&0.79\\
2452317.6479750 & 3982792&    10.3&0.95\\
2452321.8348344 & 3983555&    11.4&1.23\\
2452322.7265266 & 3984372&    9.9 &3.03\\
2452312.7412881 & 3984730&    11.4&3.5\\
2452373.6840808 & 4089983&     9.9&1.1\\
2452373.6839702 & 4090425&    12.6&1.2\\
2452373.7140655 & 4090791&    10.6&0.68\\
2452375.6392709 & 4122465&    12.8&1.0\\
2452374.7700070 & 4124076&    12.5&1.23\\
2452581.9494464 & 4134940&  12.0  &1.83\\
2452583.9095168 & 4137288&  11.5  &1.74\\
2452584.8812628 & 4144503&  13.8  &0.92\\
2452585.9821875 & 4148953&  12.0  &1.16\\
2452586.8937641 & 4146104&  12.3  &0.7\\
2452665.7845581 & 4255806&  13.7  &0.72\\
2452669.7970851 & 4256260&  10.6  &1.52\\
2452696.8561592 & 4257768&  12.8  &0.96\\
2452724.6624548 & 4267394&  10.5  &0.81\\
2453381.7442572 & 4409917&  11.5  &1.41\\
2453439.7653860 & 4433212&  13.9  &1.0\\ 
2453446.6073830 & 4435959&  15.2  &0.71\\
2453473.6191370 & 4446804&  15.1  &0.75\\ 
\hline
\end{tabular}
\end{center}
\end{table}

\begin{table}
\begin{center}
\begin{tabular}{|r|l|c|c|}  \hline
{\bf Time of Maximum} & {\bf Epoch}   & {\boldmath$(O-C)$}  & {\boldmath$\sigma$}   \\
BJDD                 &    &(sec)       & (sec)  \\ \hline
2453709.9576790&4541692&15.36&1.58\\  
2453713.7883270&4543230& 9.76&1.36\\  
2453795.7202970&4576125&13.89&1.14\\
2453798.7440310&4577339&14.88&1.56\\ 
2453800.7291150&4578136&13.82&1.32\\ 
2454090.9940570&4694675&16.41&1.92\\ 
2454097.8708770&4697436&13.67&1.26\\ 
2454175.7329770&4728697&13.56&2.10\\ 
2454505.6674710&4861163&16.64&1.09\\ 
2454536.7864160&4873657&17.32&1.16\\ 
2454538.7914220&4874462&15.95&1.21\\ 
2454539.5760330&4874777&19.16&1.43\\ 
2454860.8527290&5003767&14.62&1.18\\ 
2454861.8714540&5004176&16.72&1.16\\ 
2454865.7718960&5005742&15.80&1.43\\ 
2454935.282668&5033650&17.80&2.00\\ 
2454937.327536&5034471&17.34&2.67\\ 
2454945.716259&5037839&18.20&1.66\\ 
2455215.8113837&5146280&17.03&1.29\\ 
2455216.7429028&5146654&16.45&1.18\\  
2455243.7347524&5157491&18.16&1.26\\ 
2455266.604463&5166673&18.74&1.76\\ 
2455272.6766822&5169111&17.25&1.83\\ 
2455273.588298&5169477&17.26&1.63\\ 
2455274.592065&5169880&19.53&1.42\\ 
2455276.6069443&5170689&15.42&3.27\\ 
2455542.8289250&5277575&19.59&1.28\\ 
2455567.8558979&5287623&18.64&1.48\\ 
2455570.8223324&5288814&18.49&1.33\\ 
2455593.7542828&5298021&17.52&1.29\\
2456395.6136513&5619961&19.00&1.44\\ 
2456405.6736629&5624000&21.75&1.16\\ 
2457039.68401488&5878550&21.05&4.19\\  
2457040.73759366&5878973&21.76&1.79\\
2457095.73749568&5901055&24.57&1.27\\
2457097.63542361&5901817&25.14&0.96\\
2457100.62178721&5903016&25.28&0.62\\
2457101.62556613&5903419&27.23&14.81\\
2458189.71735867&6340279&27.19&1.32\\
2458190.59411461&6340631&29.42&1.41\\
2458191.66260081&6341060&26.95&1.22\\
2458192.59418283&6341434&31.81&1.50\\
2458192.68629503&6341471&28.00&2.65\\
2458146.67287842&6322997&25.35&0.79\\
\hline
\end{tabular}
\end{center}
\end{table}

\begin{table}
\begin{center}
\begin{tabular}{|r|l|c|c|}  \hline
{\bf Time of Maximum} & {\bf Epoch}   & {\boldmath$(O-C)$}  & {\boldmath$\sigma$}   \\
BJDD                 &    &(sec)       & (sec)  \\ \hline
2458145.70398597&6322608&24.83&0.58\\
2458144.72013874&6322213&23.40&0.85\\
2458488.81432050&6460364&26.35&0.54\\
2458870.45096380&6613588&27.75&3.14\\
2458902.63843304&6626511&29.26&13.8\\

\hline

\end{tabular}
\end{center}
\end{table}


\appendix

\section{Estimate of Phase Drift of 215~s Mode}
\label{drift}

According to \citet{mikemon20}, modes experience a small phase shift
at their outer turning point due to the changing depth of the
convection zone. If we assume that the average phase shift (due to the
presence of multiple modes) is essentially random, then we can treat
the accumulated phase after multiple reflections as a random walk.
Denoting the average phase shift after one reflection as $\langle
\Delta \phi \rangle$, then the average total phase shift after $N$ cycles is
given by
\begin{equation}
  \langle \Delta \phi \rangle_{\rm tot} = N^{1/2} \langle \Delta \phi
  \rangle \, .
\end{equation}
Values of  $\langle \Delta \phi \rangle$ can be obtained from the
damping rate $\gamma$ via the following relation \citep[equation~15 of][]{mikemon20}:
\begin{equation}
  \gamma = \frac{1}{n P} \left(1 - \frac{\sin \langle\Delta
      \phi\rangle}{\langle\Delta \phi \rangle}  \right) \approx
  \frac{1}{6 n P} \, \langle \Delta \phi
  \rangle^2,
\end{equation}
where $n$ is the radial order of the mode and $P$ is its period. 
Thus, we find that
\begin{equation}
  \langle \Delta \phi \rangle_{\rm tot} = \left( 6 n \, P \,\gamma \,N
  \right)^{1/2}.
\end{equation}

For the relevant mode in G117-B15A, the total number of cycles
is $N \approx 6 \times 10^6$, $n = 2$, $P \approx 215$~s, and, from
Fig.~9a of \citet{mikemon20}, $\gamma <
10^{-15}\,\rm s^{-1}$ (and possibly \emph{much} smaller than this), which yields a total phase shift of
  $\langle \Delta \phi \rangle_{\rm tot} \approx 4 \times
  10^{-3}\,\rm rad$. Thus, the average shift of the last point in the
  $O-C$ diagram should be $P \langle \Delta \phi \rangle_{\rm tot}/2
  \pi \approx 0.13$~s. Given this small value, the analysis of G117-B15A should
  be unaffected by the time-dependent effects of the surface
  convection zone.
  
\section{Effect of a Changing Magnetic Field}
\label{magpert}
A pulsating white dwarf with a magnetic field should have its oscillation frequencies perturbed by the field. If that field changes with time (as has been directly observed in many astronomical objects) then the oscillation frequencies will also change with time. 

For the case of oscillation frequencies perturbed by slow rotation, the use of perturbation theory is valid because the effects of rotation are everywhere small. This is not true for weak magnetic fields. Near the surface of a stellar model the gas pressure ($P_{\rm gas}$) approaches zero while the magnetic pressure does not. Thus, there is always a region in which $B^2/8 \pi > P_{\rm gas}$, and, since the magnetic field geometry can modify the angular structure of the modes in this region, a self-consistent treatment can be quite complex  \citep[e.g.,][]{Dziembowski96,Bigot00,Bigot02}.

Fortunately, we are only interested in the special case of the effect of a weak magnetic field on the frequencies of low-order g-modes in white dwarfs. These modes have outer turning points far below the region where $B^2/8 \pi \sim P_{\rm gas}$, so the perturbations to their frequencies do not strongly depend on their angular structure in the surface layers. Thus, a simple perturbative treatment as used in \citet{Jones89} should be adequate for these modes. 

Since we are only interested in order of magnitude estimates, we choose a constant field in the $\vec{z}$ direction aligned with the rotation axis. We also only consider the perturbation of $m=0$ modes; the perturbation of other $m$ values will be the same order of magnitude. Repeating the analysis in \citet{Montgomery94} for a ``G117-B15A-like'' model ($T_{\rm eff} =$12,400~K, $M_\star = 0.6 M_\odot$), we find that a magnetic field that decreases in strength from 280~G to 0~G over a time span of 46 years can produce $\Pdot \approx 5.1 \times 10^{-15}$~s/s for a mode with $k=2$ and $P = 209$~s. Furthermore, it is actually the change in $B^2$ that matters, i.e., $\Delta B^2$, so the same effect would be produced by a magnetic field the goes from 2814~G to 2800~G over the same time span.

\bibliography{kepler}{}
\bibliographystyle{aasjournal}



\end{document}